\documentclass[twocolumn,resetfootnote]{aastex701}

\newcommand{\ang}{\,\mbox{\AA}}	
\newcommand{\cratio}{\,$^{12}$C/$^{13}$C}	
\newcommand{\teff}{\,$T_{\mathrm{eff}}$}
\newcommand{\mass}{\,$M_{\odot}$}

\begin{document}

\title{Exploring the History of Stellar Mergers with Chemistry: Examining the Origins of Massive $\alpha$-Enriched Stars using Carbon Isotope Ratios}

\author[orcid=0000-0002-0475-3662,sname=Maas]{Zachary G. Maas}

\affiliation{Indiana University, Astronomy Department, 727 East Third Street, Bloomington, IN 47405, USA}
\email[show]{zmaas@iu.edu}  

\author[orcid=0000-0002-1423-2174, sname=Hawkins]{Keith Hawkins} 

\affiliation{Astronomy Dept., The University of Texas at Austin, 2515 Speedway Boulevard, Austin, TX 78712, USA}
\email{keithhawkins@utexas.edu}

\author[orcid=0000-0001-9583-0004,sname=Gerber]{Jeffrey M. Gerber}
\affiliation{Purdue University, Department of Physics and Astronomy
525 Northwestern Ave
West Lafayette, IN 47907, USA}
\email{gerber48@purdue.edu}

\author[orcid=0000-0002-3855-3060, sname=Hackshaw]{Zoe Hackshaw}
\affiliation{Astronomy Dept., The University of Texas at Austin, 2515 Speedway Boulevard, Austin, TX 78712, USA}
\email{zoehackshaw@utexas.edu}

\author[0000-0002-0900-6076,sname=Manea]{Catherine Manea}
\affiliation{Columbia Astrophysics Laboratory, Columbia University, New York, NY, 10027, USA}
\email{cm4582@columbia.edu}

\begin{abstract}

Recently discovered massive $\alpha$-enriched (MAE) stars have surface chemistry consistent with members of the older thick disk Milky Way population but high masses ($\sim$ 1.5 - 3 M$_{\odot}$) that suggest these stars are young. The origin of MAE stars is not fully understood although binary interactions are likely an important formation pathway. To better constrain the history of MAE stars, we have measured metallicities, carbon isotope ratios, and CNO abundances in 49 red clump stars and four red giants. Our sample included thin disk, thick disk, and MAE stars to best constrain different formation scenarios. We observed our sample stars using the Tull spectrograph on the McDonald 2.7m telescope and derived abundances using synthetic spectra created by the \texttt{Turbospectrum} radiative transfer code. Overall, we find that 10 of our red clump MAE stars are consistent with the average thick disk carbon isotope ratio of \cratio{} = 8.2 $\pm$ 3.4. We find five MAE stars that have significantly higher carbon isotope ratios (\cratio{} $>$ 15) than stars at similar metallicities. Two of the anomalous stars show abundance patterns consistent with AGB mass transfer while three MAE stars have \cratio{} ratios similar to massive, single RC stars and show no clear signs of binarity from radial velocity monitoring or from the Gaia RUWE measurement. Overall, we find that carbon isotope ratio measurements provide a unique constraint when discerning the possible origins of MAE stars.

\end{abstract}


\section{Introduction}
Industrial scale chemical evolution surveys have revealed new stellar populations in the Milky Way. The information from asteroseismology and spectroscopic abundances has discovered stars with high masses ($\sim$ 1.5 - 3 M$_{\odot}$) and high [$\alpha$/Fe] ratios \citep{martig15,chiappini15}. The relatively high masses imply these stars are young while the high [$\alpha$/Fe] ratios are consistent with the chemical composition of stars in the Milky Way thick disk. Multiple formation pathways have been proposed for these objects. One possibility is that MAE stars\footnote{We refer to these objects as massive $\alpha$-enriched stars instead of $\alpha$-rich young stars since the stellar ages are inferred from masses which may have changed over time due to mass-transfer events or stellar mergers.} may be born from gas with less Type Ia SN contributions and then are displaced outward by the bar. Another scenario suggests that MAE stars were older objects with enhanced masses due to stellar mergers or mass transfer \citep{martig15,chiappini15}. High-resolution optical follow-up of MAE stars have corroborated the $\alpha$-enhancement and found radial velocity variations consistent with a binary system \citep{jofre16,yong16,matsuno18}. Additional long-term study of MAE stars have also found a significant fraction of their sample appear as binary stars while some MAE stars show no radial velocity variation \citep{jofre23}.

Another potential method to exploring the origins of MAE stars is to examine elements altered through internal mixing processes during stellar evolution. Proton-capture (p-capture) burning cycles are strongly temperature-dependent and therefore will correlate to mass. As the star evolves along the sub-giant and red giant branch, the p-capture burning products created in the CNO cycle are transported from the core to the surface, and specific elements may be used as a diagnostic of nucleosynthesis occurring in the interior of the star \citep{iben67}. Ratios of $^{12}$C, $^{13}$C, and N are especially sensitive to mass, metallicity, and evolutionary state. Models of disk evolution predict a large fraction of mergers of low-mass stars occur after the first dredge-up and therefore their C and N abundances are expected to be consistent with low mass stars \citep{izzard18}. Follow-up measurements of CNO abundances in 51 MAE stars found that some stars had [C/N] ratios that reflected low-mass stars after the first dredge up while other were consistent with higher mass stars \citep{jofre16,hekker19,sun20,zhang21}. The [C/N] ratio is mass-sensitive for red-giant stars and the wide range of [C/N] found in MAE stars suggests multiple pathways of binary star evolution necessary to produce a MAE star \citep{jofre16,hekker19,sun20,zhang21}. 

 A statistical analysis of MAE stars identified from the LAMOST survey demonstrated that MAE stars had similar kinematics and [C/N] ratios as thick disk stars \citep{huang20,sun20}. Beyond [C/N] and kinematics, some MAE stars have been found to be s-process enhanced in elements like Ce and Ba, elements primarily synthesized in asymptotic giant branch stars \citep{zhang21,cerqui23}. Multiple binary interaction scenarios have been discussed in the literature and the possible components include combinations of main-sequence, red giant branch stars, and asymptotic giant branch stars that can completely merge, have a common envelop phase, or undergo mass-transfer \citep{jofre16,zhang21,cerqui23}. Similar pathways have been necessary when discerning the history of possible field blue straggler stars.

One potentially important way to distinguish the formation scenarios of evolved MAE stars is through measurements of the \cratio{} ratio. Since the CNO-process directly increases the $^{13}$C abundance in a star's interior, stellar mixing changes the surface abundance beginning with the first dredge-up on the sub giant branch when $^{13}$C rich material is brought from the interior to the surface \citep{iben65,charbonnel95}. The mixing efficiency varies depending on metallicity; red giant branch metal rich disk stars have been observed to have carbon isotope ratios of 20-40, while the decrease is even more drastic for metal poor stars \citep{boothroyd99} (as a reference main-sequence solar-twins span 70.9 $<$ \cratio{} $<$ 103.4, \citealt{botelho20}).

Non-canonical extra-mixing occurs for low-mass stars between 0.5 \mass{} $\lesssim$ M$_{star}$ $\lesssim$ 2.0 \mass{} after the star passes the luminosity function (LF) bump on the red giant branch. The extra-mixing further lowers the carbon isotope ratios as additional CNO-cycle material is dredged up from the H-burning shell around the core to the photosphere. Multiple theoretical models have been purposed to explain the extra-mixing, such as thermohaline mixing \citep{kippenhahn80,charbonnel07,eggleton08,cantiello10,wachlin11,lagarde12}, rotation-induced mixing \citep{sweigart79,chaname05,palacios06}, magnetic mixing mechanisms \citep{busso07,nordhaus08,palerini09}, and internal gravity waves \citep{denissenkov00}. Observational evidence of extra-mixing has been identified in the CNO abundances of field giants and clusters. Specifically, low \cratio{} ratios have been observed in field stars \citep{gratton00,keller01,spite06,tautvaisien13,aguileragomez23}, open clusters \citep{smiljanic09,tautvaisien16,szigeti18,mccormick23}, and in globular clusters \citep{suntzeff91,briley97,shetrone03,smith07,recioblanco07,maas19}. 

The summary of these observations and models demonstrates that the extra-mixing processes are dependent on stellar mass and metallicity. For the \cratio{} ratio specifically, more metal-poor stars dredge up more $^{13}$C and low-mass stars dredge up more $^{13}$C as predicted by thermohaline mixing models (e.g. \citealt{charbonnel10,lagarde12}) and from observations of stars (e.g. \citealt{tautvaisien16,aguileragomez23}). Specifically, higher mass core He-burning stars are expected to have \cratio{} ratios of $\sim$ 20, while lower mass core He-burning stars have \cratio{} ratios of $\sim$ 10 \citep{charbonnel10,lagarde12}.

One intriguing avenue to study MAE stars is by measuring the \cratio{} ratio and comparing to disk stars at similar metallicities. The types and timing of mass transfer and/or merger events may be constrained by this photospheric abundance since there are clear points in stellar evolution where the \cratio{} ratio changes. To this end, we have measured the \cratio{} in multiple thin disk, thick disk, and MAE stars. The observations and sample selection are described in section \ref{sec:sample_observations}. The abundance measurement methodology is detailed in section \ref{sec:abund_method} with a description of the measurement uncertainty in section \ref{sec:uncer}. The discussion of the results is found in section \ref{sec:discussion}. Finally, the results are described in \ref{sec:conclusion}.

\section{Sample Selection and Observations}
\label{sec:sample_observations}
\subsection{Selecting Stars in Different Stellar Populations}
\label{subsec:sample}
In order to detect differences in the composition of MAES stars and typical field giants, we require a selection of stars at similar evolutionary states to minimize systematic differences due to stellar mixing, a sample of stars at different masses, and finally, stars that are bright enough to achieve the required signal-to-noise ratios (S/N) to detect weak molecular absorption lines. Our sample has been selected from the \citet{huang20} catalog of $\sim$ 140,000 red clump (RC) stars. The RC stars were identified using data from the Large Sky Area Multi-Object Fibre Spectroscopic Telescope (LAMOST) survey and the methodology of \citet{huang15}. Masses and ages were derived for the RC sample using a KPCA \citep{scholkopf98} machine-learning analysis. The test and training samples were created from a crossmatch of 4185 stars with measured astereoseismic masses and ages from \citet{yu18}. The principle component analysis was applied to the LAMOST blue-arm spectrum for each object, which spanned from 3900 $\mbox{\AA}$ to 5500 $\mbox{\AA}$. No systematic discrepancies between age and mass residuals with atmospheric parameters were detected \citep{huang20}. Stellar population mass distributions from \citet{sun20} are similar to RGB stars from \citep{zhang21}. We note that some low-mass ($<$ 1.3 $M_{\odot}$) MAE stars exist in our sample and are slightly offset compared to asteroseismic MAE star samples (e.g. \citealt{martig15}) but within the typical errors for mass measurements ($\pm$ 0.2 M$_{\odot}$).

The mass measurement methodology follows previous KPCA analysis of LAMOST spectra \citep{xiang17,wu18,wu19} and was validated using observations of open cluster members. A combination of spectral features sensitive to atmospheric parameters, [$\alpha$/Fe], and CN abundances all are potentially used in the KPCA analysis. \citet{sun20} states that all features, not just CN, were used when determining masses and \citet{wu18} shows that ages from only [C/N] ratios for red giant branch stars are correlated, but not identical, to ages from KPCA analysis.

We selected thin disk, thick disk, and MAES stars from this RC catalog for observations. Selection criteria was adopted from \citet{sun20} and included:

\begin{itemize}
\item{Quality Cuts: Age $<$ 14 Gyr, Age error less than 50$\%$, [Fe/H] $>$ --1~dex, 10 $\%$ distance errors, and Gaia DR3 Gmag $<$ 12 mag}
\item{Thin Disk: 0 $<$ Age $<$ 6 Gyr, [$\alpha$/Fe] $<$ 0.1 dex}
\item{Thick Disk: 8 $<$ Age $<$ 14 Gyr, [$\alpha$/Fe] $>$ 0.2 dex}
\item{MAES: 0 $<$ Age $<$ 6 Gyr, [$\alpha$/Fe] $>$ 0.2 dex}
\end{itemize}

\begin{figure*}

	\includegraphics[width=7in]{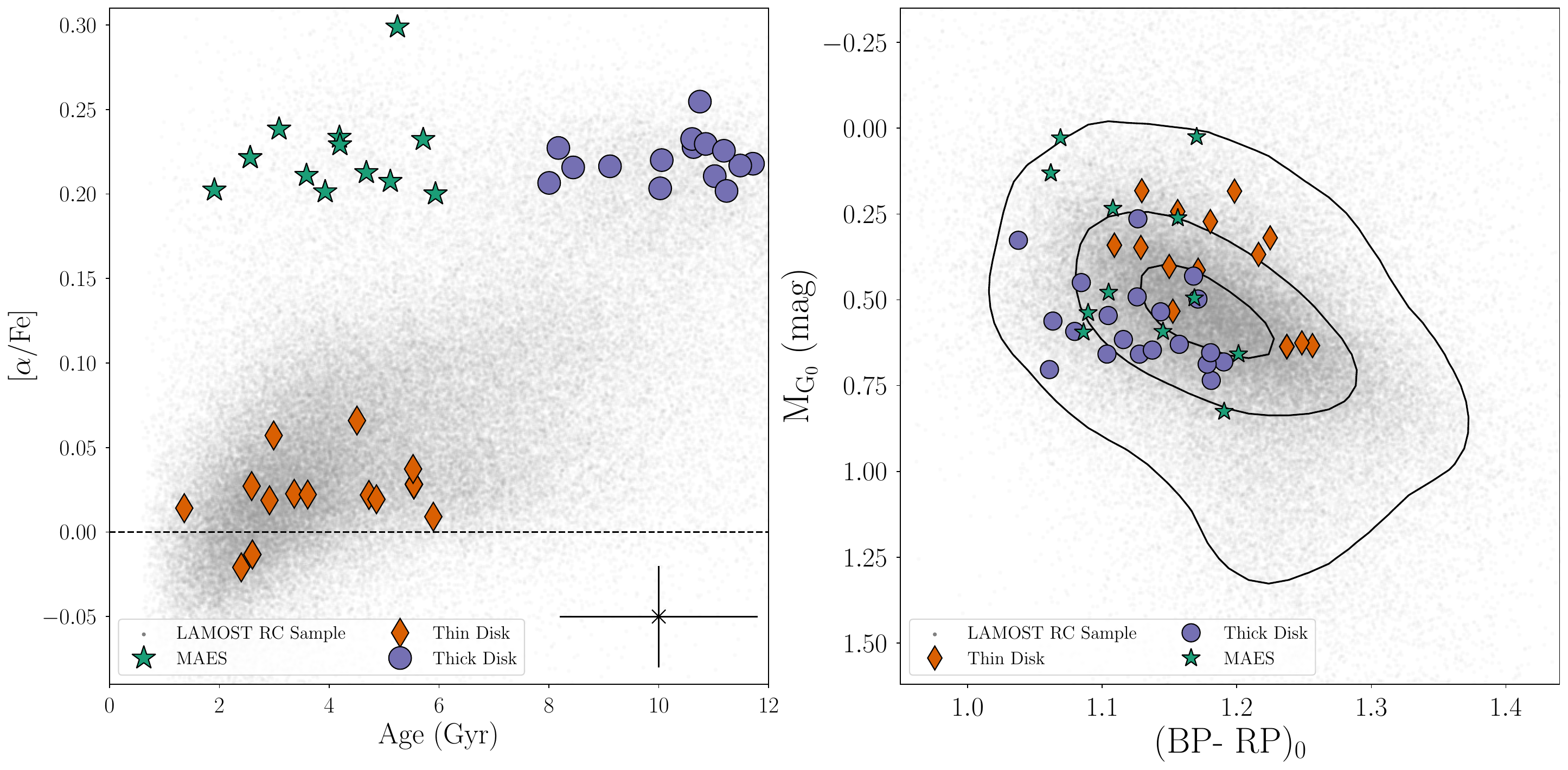}
    \caption{Left: The stars selected from \citet{huang20} and analyzed in this work are represented as green stars for MAE stars, blue circles for thick disk stars, and orange diamonds for thin disk stars. Gray points are the rest of the LAMOST RC catalog that was unobserved. The figure shows ages compared to [$\alpha$/Fe] enrichment with parameters from \citet{huang20}. The mean age and [$\alpha$/Fe] uncertainties for our sample stars is shown as a black x. Right: We show a color-magnitude diagram for our targets selected from \citet{huang20}. Contours represent 16$\%$, 50$\%$, 84$\%$ density values for the total sample. The extinction correction methodology is described in subsection \ref{subsec:sample}.}
    \label{fig:sample_abun_phot}
\end{figure*}

We added the G-band magnitude restriction from Gaia DR3 data \citep{gaiadr3} to ensure we can achieve our S/N target of $\sim$ 100 at 8000 $\ang$ (the wavelength region near the CN feature from which we determine \cratio{} ratios). The total RC sample that meets these criteria are plotted in Fig. \ref{fig:sample_abun_phot} and are color-coded depending on stellar population membership given in Table \ref{tab:observing_log}. 

We also wanted to remove any potential non-RC star contaminants drawn to ensure outlier stars do not impact our analysis of MAE star origins. We derived absolute G band magnitudes to see if our stars were photometrically consistent with RC stars using Gaia DR3 photometry \citep{gaia21,gaiadr3}, extinction corrections from \texttt{dustmaps} (v. 2019) \citep{green19}, and dust extinction conversions to Gaia filters from \citet{wang19}. The ``bayestar2019" dust map option was used to determined the extinction for each star. The distances were adopted from the \citet{bailerjones21} catalog to calculate the absolute G magnitude for each star in our sample. We plot our stars along with the targets with a subsample of \citet{huang20} stars in Fig. \ref{fig:sample_abun_phot}. We also only use stars that meet the photometric criteria of: 0.0 $<$ M$_{G}$ $<$ 0.9 and 1.02 $<$ BP-RP $<$ 1.27. We adopt these limits because redder stars are more likely to be red giants with high \cratio{} ratios, brighter stars may be beginning to ascend the AGB and start dredging up $^{12}$C, and finally low-luminosity stars may be anomalous objects, such as pre-main sequence stars. All of our stars fall within or near the 84$\%$ percentile of stars with similar photometric properties to help remove outliers, as seen in Fig. \ref{fig:sample_abun_phot}. 

We supplement our RC sample chosen from \citet{huang20} with additional MAE stars that were identified from a combination of asteroseismology and chemical composition analysis. We separately identify the stars in our sample with asteroseismic masses as Kepler stars in plots and tables. Information about these targets is described in Table \ref{tab:observing_log_kic}. The objects were identified by \citet{hekker19,jofre23}, with masses originally derived from APOKASC \citep{pinsonneault18}.

\subsection{Observation Summary and Data Reduction}

Observations were obtained using the Tull Coud\'e spectrograph on the McDonald 2.7m telescope. We used the TS23 set-up with slit 4 to achieve a resolution of R $\sim$ 60,000. The $^{13}$CN feature at $\sim$ 8005 $\ang$ was targeted and the CCD prism/grating settings were adjusted to ensure this feature was not off or at the edge of the CCD. We also observed RC stars from \citet{afsar12} to validate our methodology since these objects were previously analyzed using the same instrument/facility combination and have derived carbon isotope ratios using the same CN feature used in this work. 

Data reduction was done using IRAF\footnote{IRAF is distributed by the National Optical Astronomy Observatory, which is operated by the Association of Universities for Research in Astronomy, Inc., under cooperative agreement with the National Science Foundation} \citep{tody86,tody93} with the PyRAF\footnote{PyRAF is a product of the Space Telescope Science Institute, which is operated by AURA for NASA.} wrapper \citep{pyraf12}. The science images were trimmed and bias corrected with \texttt{ccdproc}. Cosmic rays were removed using the \texttt{cosmicrays} task. Pixel-to-pixel variations were corrected with a quartz lamp flat-field image. The echelle spectra were corrected for scattered light using \texttt{apscatter} and extracted with \texttt{apall}. 

The wavelength solution for the observations was determined with ThAr lamp observations for each night and the dispersion to the emission line location fits were typically 0.003 \ang. The reduced spectra were then normalized using a spline function and the separate apertures were combined. Telluric standards were observed and the CN region was examined for potential contamination. The \texttt{telluric} task in IRAF was used to removed telluric lines for stars where the $^{13}$CN line region when a blend was detected.

We used \texttt{iSpec} \citep{blanco-cauresma14,blanco-cauresma19} to measure and correct for the radial velocities of each star using a cross-correlation technique with a solar template spectrum. Typical radial velocity uncertainties from \texttt{iSpec} cross-correlation range for 0.1 km/s to 0.2 km/s. When compared to Gaia DR3 \citep{gaiadr3}, our sample has an average radial velocity difference of --0.04 $\pm$ 2.46 km/s. There are four outlier stars with radial velocity differences beyond 5 km/s and when they are removed the average becomes 0.05 $\pm$ 1.21 km/s. Finally, stars with multiple exposures were mean combined after the RV correction. Due to significant blending and low S/N ratios in the blue region of the spectrum, we restricted our analysis to 4700 $\ang$ - 8800 $\ang$ for abundance analysis. Observation dates and signal to noise ratios for our sample are listed in Table \ref{tab:observing_log}. 

\begin{table*}
	\begin{center}
    \caption{LAMOST RC Observation Log}
	\label{tab:observing_log}
	\begin{tabular}{lcccccccccr} 
		\hline
		Gaia DR3 ID & Date Obs. & S/N Ratio & G & A$_{\mathrm{G}}$ & M$_{\mathrm{G}}$ & Age & $\sigma$ Age & [$\alpha$/Fe] & $\sigma$ [$\alpha$/Fe] & Pop.\\
		& (UT Date) & pixel$^{-1}$ at 8000 $\ang$ & (mag) & (mag) & (mag) & (Gyr) & (Gyr) & (dex) & (dex) & \\
		\hline
1520951449701176832 & 2021-01-22 & 159 & 9.45 & 0.01 & 0.50 & 11.72 & 3.18 & 0.22 & 0.03 & ThickDisk \\
3872367429179326592
& 2021-01-22 & 117 & 10.23 & 0.01 & 0.23 & 4.20 & 1.14 & 0.23 & 0.03 & MAES \\
3283445774160853632 & 2021-01-23 & 117 & 10.30 & 0.71 & 0.54 & 12.34 & 3.35 & 0.21 & 0.03 & ThickDisk \\
3876280792926214272 & 2021-01-23 & 108 & 10.67 & 0.13 & 0.59 & 3.59 & 0.98 & 0.21 & 0.03 & MAES \\
723448831119162752 & 2021-01-23 & 147 & 9.90 & 0.01 & 0.66 & 11.02 & 2.99 & 0.21 & 0.03 & ThickDisk \\
3222153253655970176 & 2021-01-23 & 120 & 10.08 & 0.15 & 0.55 & 13.07 & 3.54 & 0.21 & 0.03 & ThickDisk \\
672933143771515392 & 2021-01-23 & 105 & 10.42 & 0.10 & 0.54 & 5.24 & 1.42 & 0.30 & 0.03 & MAES \\
3813076676968705408 & 2021-01-23 & 208 & 9.92 & 0.01 & 0.63 & 12.69 & 3.44 & 0.22 & 0.03 & ThickDisk \\
3878388801529524480 & 2021-02-21 & 93 & 10.63 & 0.02 & 0.13 & 2.56 & 0.70 & 0.22 & 0.03 & MAES \\
4008867227922925824 & 2021-02-21 & 186 & 9.33 & 0.04 & 0.73 & 8.00 & 2.17 & 0.21 & 0.03 & ThickDisk \\
607150428232731008 & 2021-02-21 & 123 & 9.50 & 0.16 & 0.66 & 5.11 & 1.39 & 0.21 & 0.03 & MAES \\
3163974593033517824 & 2021-02-21 & 91 & 10.16 & 0.02 & 0.69 & 9.11 & 2.47 & 0.22 & 0.03 & ThickDisk \\
971925848973504512 & 2021-02-21 & 119 & 9.68 & 0.56 & 0.03 & 4.19 & 1.15 & 0.23 & 0.03 & MAES \\
1235499505072865792 & 2021-02-22 & 138 & 9.53 & 0.04 & 0.59 & 5.93 & 1.62 & 0.20 & 0.03 & MAES \\
636614659798851712 & 2021-02-22 & 150 & 9.93 & 0.05 & 0.03 & 3.09 & 0.84 & 0.24 & 0.03 & MAES \\
3882249324293741440 & 2021-02-22 & 133 & 10.39 & 0.07 & 0.49 & 10.61 & 2.88 & 0.23 & 0.03 & ThickDisk \\
3326023296796724096 & 2021-02-22 & 78 & 9.39 & 0.10 & 0.63 & 5.90 & 1.60 & 0.01 & 0.03 & ThinDisk \\
3130104824535988096 & 2021-02-22 & 153 & 10.16 & 0.29 & 0.59 & 10.75 & 2.91 & 0.25 & 0.03 & ThickDisk \\
1486503509885506944 & 2021-04-25 & 106 & 8.86 & 0.02 & 0.24 & 4.51 & 1.22 & 0.07 & 0.03 & ThinDisk \\
1330490025449104640 & 2021-04-25 & 96 & 9.44 & 0.12 & 0.53 & 4.86 & 1.32 & 0.02 & 0.03 & ThinDisk \\
1312079562236150656 & 2021-04-25 & 91 & 9.25 & 0.04 & 0.41 & 2.59 & 0.71 & 0.03 & 0.03 & ThinDisk \\
1388620891675799168 & 2021-04-25 & 127 & 7.46 & 0.02 & 0.40 & 2.91 & 0.79 & 0.02 & 0.03 & ThinDisk \\
3833688568619253120 & 2021-04-27 & 81 & 8.83 & 0.02 & 0.63 & 3.61 & 0.99 & 0.02 & 0.03 & ThinDisk \\
3976085529258764800 & 2022-03-11 & 78 & 9.27 & 0.04 & 0.64 & 5.54 & 1.50 & 0.03 & 0.03 & ThinDisk \\
688166190179213184 & 2022-03-11 & 64 & 9.40 & 0.20 & 0.34 & 2.60 & 0.71 & -0.01 & 0.03 & ThinDisk \\
3811610203335180416 & 2022-03-12 & 123 & 9.80 & 0.05 & 0.70 & 10.05 & 2.73 & 0.22 & 0.03 & ThickDisk \\
866507457241940480 & 2022-03-12 & 105 & 8.85 & 0.12 & 0.35 & 2.40 & 0.65 & -0.02 & 0.03 & ThinDisk \\
638939267896892800 & 2022-03-12 & 67 & 10.19 & 0.13 & 0.26 & 1.91 & 0.52 & 0.20 & 0.03 & MAES \\
1156535637983109376 & 2022-03-12 & 157 & 9.89 & 0.09 & 0.68 & 10.03 & 2.72 & 0.20 & 0.03 & ThickDisk \\
3952269007889690752 & 2022-03-12 & 104 & 9.92 & 0.05 & 0.45 & 8.44 & 2.29 & 0.22 & 0.03 & ThickDisk \\
3976449776845278592 & 2022-03-13 & 99 & 9.11 & 0.05 & 0.18 & 2.99 & 0.81 & 0.06 & 0.03 & ThinDisk \\
685849244302001408 & 2022-03-13 & 166 & 10.21 & 0.09 & 0.48 & 5.71 & 1.55 & 0.23 & 0.03 & MAES \\
1447162502806921472 & 2022-03-13 & 111 & 9.44 & 0.04 & 0.37 & 5.53 & 1.50 & 0.04 & 0.03 & ThinDisk \\
1457380298724918528 & 2022-03-13 & 164 & 10.02 & 0.03 & 0.56 & 8.17 & 2.22 & 0.23 & 0.03 & ThickDisk \\
4009353280781776384 & 2022-03-13 & 152 & 9.24 & 0.05 & 0.18 & 3.36 & 0.91 & 0.02 & 0.03 & ThinDisk \\
1170340693664424192 & 2022-03-13 & 102 & 10.10 & 0.07 & 0.65 & 12.52 & 3.41 & 0.24 & 0.05 & ThickDisk \\
3096606828405802240 & 2022-03-14 & 79 & 10.63 & 0.06 & 0.43 & 14.41 & 3.90 & 0.22 & 0.03 & ThickDisk \\
1500728407210595200 & 2022-03-14 & 97 & 10.16 & 0.10 & 0.26 & 11.48 & 3.11 & 0.22 & 0.03 & ThickDisk \\
4034411734894241024 & 2022-03-14 & 121 & 10.50 & 0.05 & 0.66 & 11.19 & 3.03 & 0.23 & 0.03 & ThickDisk \\
842398878096774784 & 2022-03-14 & 108 & 8.76 & 0.03 & 0.32 & 4.72 & 1.29 & 0.02 & 0.03 & ThinDisk \\
579491624824868608 & 2022-03-14 & 121 & 10.46 & 0.05 & 0.83 & 3.93 & 1.07 & 0.20 & 0.03 & MAES \\
3138549726673657600 & 2022-03-15 & 95 & 9.39 & 0.02 & 0.27 & 1.36 & 0.38 & 0.01 & 0.03 & ThinDisk \\
640282939825239936 & 2022-03-15 & 128 & 9.94 & 0.05 & 0.62 & 10.63 & 2.88 & 0.23 & 0.03 & ThickDisk \\
3671355056986858880 & 2022-03-15 & 70 & 10.75 & 0.09 & 0.49 & 4.68 & 1.27 & 0.21 & 0.03 & MAES \\
3869545635666293504 & 2022-03-15 & 150 & 10.69 & 0.06 & 0.65 & 11.23 & 3.05 & 0.20 & 0.03 & ThickDisk \\
3973325544619220224 & 2022-03-15 & 131 & 10.17 & 0.34 & 0.33 & 10.85 & 2.94 & 0.23 & 0.03 & ThickDisk \\
		\hline
	\end{tabular}
\end{center}
G-band magnitudes are from Gaia eDR3 \citet{gaia21}, ages and [$\alpha$/Fe] are from \citet{huang20}. The "Pop." column refers to the assigned stellar population and includes: TD =  Thin Disk, ThD = Thick Disk, and MAES = Massive $\alpha$-Enhanced Star.
\end{table*}

\begin{table}
	\begin{center}
    \scriptsize
    \caption{Kepler MAES Observation Log}
    \label{tab:observing_log_kic}
	\begin{tabular}{lcccccccccr} 
		\hline
		Kepler ID & Date Obs. & S/N Ratio & G & Spectral \\
		& (UT Date) & pixel$^{-1}$ at 8000 $\ang$ & (mag) & Type & \\
		\hline
KIC 8539201  & 2021-04-25 & 123 & 9.45 & RC \\
KIC 3833399  & 2022-03-12 & 93 & 9.45 & RC \\
KIC 5687374 & 2023-07-11 & 125 & 10.42 & RG \\
KIC 5966873 & 2023-07-11 & 85 & 9.75 & RG \\
KIC 3455760  & 2023-07-12 & 89 & 10.9 & RG \\
KIC 10525475 & 2023-07-12 & 87 & 10.73 & RC \\
KIC 4149831  & 2023-07-13 & 90 & 10.49 & RC \\
		\hline
	\end{tabular}
    \end{center}
   G-band magnitudes of Kepler stars in our sample are from Gaia DR3 \citet{gaiadr3}. RC = Red clump and RG = Red giant for spectral type. 
\end{table}

\section{Abundance Measurement Methodology}
\label{sec:abund_method}

\subsection{Atmospheric Parameter Measurements with \texttt{BACCHUS}}
\label{subsec:bacchus}
Our primary goal is to measure precise carbon isotope ratios for our sample of stars using a $^{13}$CN absorption line at 8005 \ang{}. We require the full CNO abundances, since the molecular equilibrium and therefore CN molecular feature line strengths, depend on the chemical composition of the star. Our measurements were performed using \texttt{turbospectrum} \citep{plez12,gerber22}, MARCS model atmospheres \citep{gustafsson08}, and the Gaia-ESO Survey linelist \citep{heiter21} which uses CN line data from \citet{brooke14}. All abundances reported are scaled to the solar abundances of \citet{magg22}.

We derived atmospheric parameters using the Brussels Automatic Code for Characterizing High accUracy Spectra (BACCHUS) \citep{bacchus16}. The BACCHUS pipeline \citep{masseron16} was utilized to measure equivalent widths of Fe I and Fe II lines and iteratively measured atmospheric parameters through iron excitation-ionization balance. BACCHUS outputs include the effective temperature, surface gravity, [Fe/H], microturbulence, and convolution parameter used to smooth the spectra to compensate for the instrument profile, stellar broadening mechanisms, etc. We additionally measured C and Mg abundances with BACCHUS by analyzing atomic C I lines (near $\sim$ 5,000 $\ang$) and Mg I features. The final atmospheric parameters are listed in Table \ref{tab:atmoparams}.  

\begin{table}
	\begin{center}
    \scriptsize
    \caption{Star Atmospheric Parameters}
	\label{tab:atmoparams}
	\begin{tabular}{lccccccr} 
		\hline
		Gaia DR3 ID & Kepler ID &  \teff{} &  log(g) & [Fe/H] & $\xi$ \\ &  & (K) & (dex)  &  (dex)  & (kms$^{-1}$) \\
		\hline
1520951449701176832 & - &4792 & 2.49 & -0.45 & 1.48 \\
3872367429179326592 & - &4964 & 2.68 & -0.58 & 1.46 \\
3283445774160853632 & - &4844 & 2.10 & -0.74 & 1.52 \\
3876280792926214272 & - &4923 & 2.27 & -0.50 & 1.34 \\
723448831119162752 & - &4933 & 2.62 & -0.59 & 1.53 \\
3222153253655970176 & - &4952 & 2.59 & -0.55 & 1.51 \\
672933143771515392 & - &4916 & 2.12 & -0.87 & 1.62 \\
3813076676968705408 & - &4833 & 2.46 & -0.49 & 1.44 \\
3878388801529524480 & - &5033 & 2.62 & -0.75 & 1.39 \\
4008867227922925824 & - &4781 & 2.46 & -0.39 & 1.56 \\
607150428232731008 & - &4546 & 2.30 & -0.51 & 1.25 \\
3163974593033517824 & - &4839 & 2.27 & -0.46 & 1.39 \\
971925848973504512 & - &4819 & 2.59 & -0.74 & 1.35 \\
1235499505072865792 & - &4796 & 2.25 & -0.44 & 1.38 \\
636614659798851712 & - &4985 & 2.44 & -0.60 & 1.35 \\
3882249324293741440 & - &4848 & 2.48 & -0.58 & 1.56 \\
3326023296796724096 & - &4695 & 3.07 & 0.24 & 1.35 \\
3130104824535988096 & - &4868 & 2.62 & -0.49 & 1.43 \\
1486503509885506944 & - &4779 & 2.44 & -0.34 & 1.24 \\
1330490025449104640 & - &4742 & 2.58 & -0.12 & 1.32 \\
1312079562236150656 & - &4784 & 2.57 & -0.03 & 1.26 \\
1388620891675799168 & - &4857 & 2.80 & -0.16 & 1.34 \\
3833688568619253120 & - &4616 & 2.69 & 0.14 & 1.27 \\
3976085529258764800 & - &4708 & 2.77 & 0.07 & 1.36 \\
688166190179213184 & - &4644 & 2.77 & 0.11 & 1.30 \\
3811610203335180416 & - &5128 & 2.58 & -0.64 & 1.64 \\
866507457241940480 & - &4773 & 2.72 & -0.02 & 1.28 \\
638939267896892800 & - &4980 & 2.93 & -0.38 & 1.57 \\
1156535637983109376 & - &4744 & 2.53 & -0.37 & 1.36 \\
3952269007889690752 & - &4960 & 2.62 & -0.62 & 1.53 \\
3976449776845278592 & - &4716 & 2.46 & -0.09 & 1.46 \\
685849244302001408 & - &4938 & 2.73 & -0.49 & 1.46 \\
1447162502806921472 & - &4656 & 2.27 & -0.14 & 1.39 \\
1457380298724918528 & - &4983 & 2.51 & -0.70 & 1.37 \\
4009353280781776384 & - &4852 & 2.70 & -0.11 & 1.51 \\
1170340693664424192 & - &4756 & 2.55 & -0.37 & 1.40 \\
3096606828405802240 & - &4792 & 2.65 & -0.41 & 1.43 \\
1500728407210595200 & - &4773 & 2.50 & -0.53 & 1.51 \\
4034411734894241024 & - &4741 & 2.67 & -0.30 & 1.40 \\
842398878096774784 & - &4674 & 2.44 & -0.05 & 1.28 \\
579491624824868608 & - &4824 & 3.00 & -0.27 & 1.43 \\
3138549726673657600 & - &4837 & 2.83 & 0.03 & 1.32 \\
640282939825239936 & - &4879 & 2.58 & -0.56 & 1.49 \\
3671355056986858880 & - &4690 & 2.78 & -0.32 & 1.21 \\
3869545635666293504 & - &4877 & 2.60 & -0.43 & 1.52 \\
3973325544619220224 & - &4807 & 2.33 & -0.65 & 1.60 \\
2117411734401094528 & KIC 8539201 & 4929 & 2.37 & -0.68 & 1.42 \\
2100234510918046592 & KIC 3833399 & 4572 & 2.33 & -0.07 & 1.26 \\
2077396108227657344 & KIC3455760 & 4581 & 2.35 & -0.16 & 1.19 \\
2100961185027552384 & KIC4149831 & 4642 & 1.92 & -0.50 & 1.56 \\
2103822354798961280 & KIC5687374 & 4259 & 1.47 & -0.57 & 1.44 \\
2077396108227657344 & KIC5966873 & 4532 & 1.96 & -0.49 & 1.41 \\
2130894220860474752 & KIC10525475 & 4748 & 2.65 & -0.23 & 1.52 \\
		\hline
	\end{tabular}
\end{center}
Solar Abundance Scale from \citet{magg22}
\end{table}

\subsection{$^{13}$C, N, and O Abundances with \texttt{TSFitPy}}

Oxygen abundance measurements were more complicated as the the 6300 \ang{} feature fell into a order gap in our observations and the 6363 \ang{} line suffers from blends with CN lines and a Ca auto-ionization line \citep{mitchell65,lambert78}. We derived oxygen abundances using the high-excitation triplet at $\sim$ 7770 $\ang$ using \texttt{TSFitPy} \citep{gerber22,storm23}. These lines are known to be significantly affected by Non-LTE affects (e.g. \citealt{magg22}) and the \texttt{TSFitPy} spectral modeling routine utilizes an NLTE version of \texttt{turbospectrum} \citep{gerber22}.  The three O I lines were fit individually and the mean abundances are reported as the [O/H] abundances for each star.

The N abundances were determined using $^{12}$CN lines near $\sim$ 8000 - 8050 $\ang$. The set of lines in the region were fit simultaneously using the \texttt{TSFitPy} routine, and the abundance that best minimized the $\chi^{2}$ was adopted as the final [N/Fe] abundance. Once all CNO abundances were derived, grids of synthetic spectra near the $^{13}$CN feature at 8005 $\ang$ were created using \texttt{TSFitPy} for each star in our sample. The \cratio{} ratios were varied by 1 in each grid, spanning from 2 $<$ \cratio{} $<$ 50. The features were fit to each observed spectrum and the model that minimized the $\chi^2$ was adopted as the final \cratio{} ratio. 

\section{Abundance Uncertainty and Literature Comparisons}
\label{sec:uncer}

\begin{figure*}[t!]
\centering
	\includegraphics[width=5in]{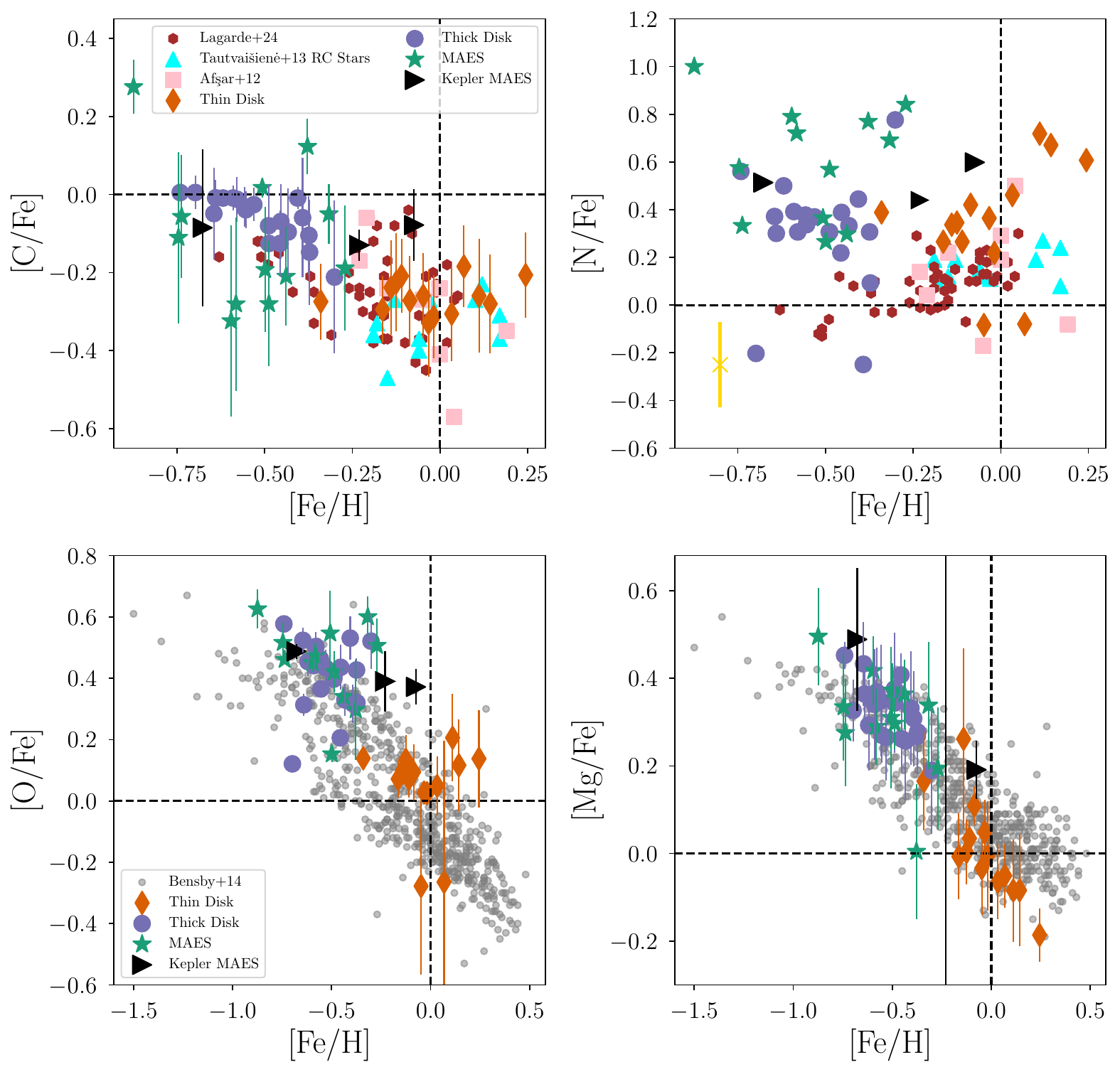}
    \caption{C, N, O, and Mg abundances are plotted for the RC stars in our sample. Representative error bars for [N/Fe] measurements are shown in the gold cross. Most symbols are the same as Fig. \ref{fig:sample_abun_phot} with the addition of Kepler MAES targets added as black triangles. The top two panels include [C/Fe] ratios and [N/Fe] ratios of probable RC stars from \citet{afsar12} as pink squares, \citet{lagarde24} as brown hexagons, and \citet{tautvaisien13} as cyan triangles. The bottom two panels include O and Mg abundances from \citet{bensby14}, represented as gray circles. Red giant abundances are discussed in section \ref{subsec:red_giant}.}
    \label{fig:cnomg_abund}
\end{figure*}

\subsection{Comparison to Af{\c{s}}ar et al. 2012}
\label{sec:comp_to_afsar12}
We observed RC abundance standard stars from \citet{afsar12} over multiple nights to examine the accuracy and precision of our atmospheric parameters and chemical abundances. Abundance standards were chosen from \citet{afsar12} since that analysis also used the Tull Spectograph on the McDonald 2.7m telescope, at a similar resolution, and derived $^{13}$C, N, and O abundances. A log of the observations, our derived parameters, and the literature abundances are listed in Table \ref{tab:abundstandards}.

Atmospheric parameters in \citet{afsar12} were derived using iron excitation abundance analysis, similar to the BACCHUS analysis in this work, with estimated uncertainties on the effective temperature of $\pm$ 150 K, the log(g) value of $\pm$ 0.3 dex, and an uncertainty of 0.1 - 0.2 dex for [Fe/H]. \citet{afsar12} also derived  solar abundances using a high-resolution solar atlas. Solar abundances were measured for different atomic and molecular lines; for example the neutral C I solar abundance was A(C) = 8.53, while the CH molecule provided a A(C) = 8.38. Since the solar scale appears to remove systematic effects in the abundance analysis we did not re-normalize the \citet{afsar12} abundances. Additionally, the common elements between both \citet{afsar12} and this study, such as the C I lines, have similar solar abundances as our adopted values from \citet{magg22}.

We determined that our internal parameter consistency is on the order of 20-30 K for \teff{}, 0.1 - 0.2 for log(g), and 0.02-0.05 dex for [Fe/H] from our repeated observations. The C and N abundances show limited star to star scatter, while the oxygen abundances have larger internal dispersion at ~0.1 dex. The abundances derived in each repeated observation are consistent with one another, as shown in Table \ref{tab:abundstandards}. When compared to \citet{afsar12}, we find consistent \teff{} but different log(g) values on the order of 0.1 - 0.3 dex and [Fe/H] of 0.05 - 0.2 dex. The carbon abundances span from a few hundredths of a dex to --0.14 dex offset for BD+27 2057. Nitrogen abundances are consistent for HIP 71837 and BD+27 2057, although HIP 56194 is offset by $\sim$0.14 dex. We find lower O abundances, possibly due to different methodologies used (e.g. the NLTE O triplet compared to the [O I] 6300 \ang{} feature). 

Overall, systematic differences are near the 1$\sigma$ errors of \citet{afsar12} and likely are due to differences in line selection, line-lists, and model atmospheres. The most discrepant parameter is log(g), but that varies in the literature for the three standard stars. \citet{wang11} found a \teff{} of 4900 K and a log(g) of 2.78 dex for HIP 56194, a log(g) of 2.90 dex derived for HIP 71837 \citep{hekker07} and a log(g) of 2.43 for BD+27 2057 from APOGEE \citep{jonsson20}. 

The carbon isotope ratios are slightly higher than those found in \citet{afsar12} for HIP 71837 and HIP 56194, however neither \cratio{} ratio measurement is more than 2$\sigma$ different. The likely issue is continuum placement for such weak lines. Importantly, the two stars (HIP 56194 and HIP 71837) with high \cratio{} ratio measurements are easily distinguished compared to the low \cratio{} ratio star BD+27 2057. 

\begin{table*}
	\begin{center}
    \scriptsize
    \caption{Standard Star Parameters and Abundances}
	\label{tab:abundstandards}
	\begin{tabular}{lcccccccccr} 
		\hline
		Name & Date Observed & SNR & T$_{\mathrm{eff}}$ & log(g) &  [Fe/H] &  $\xi$ &  [C/H] & [N/H] & [O/H]  & \cratio  \\
		&  & per pixel & (K) &  (dex)  & (dex) & (kms$^{-1}$) \\
		\hline
        From \citet{afsar12} & & & & & & & & & & \\
        \hline
        HIP 56194 & - & - & 4970 & 2.70 & 0.15 & 1.20 & --0.33 & 0.50 & 0.03 & 20 +7/-5  \\
        HIP 71837 & - & - & 4900 & 2.60 & --0.15 & 1.10 & --0.57 & 0.29 & 0.00 & 15 $\pm$ 3 \\
        BD+27 2057 & - & - & 4810 & 2.25 & --0.51 & 1.25 & --0.83 & --0.44 & --0.16 & 5 +2/-1 \\
        \hline
        This Work & & & & & & & & & & \\
        \hline
        HIP 56194 & 2021 Jan 23 & 108 & 5000 & 2.88 & --0.03 & 1.36 & --0.34 & 0.37 & --0.01 &  23 $\pm$ 4 \\
        HIP 56194 & 2021 Feb 21 & 108 & 5010 & 2.99  & --0.03 & 1.37 &  --0.30 & 0.36 & --0.03 &  28 $\pm$ 5 \\
        HIP 56194 & 2022 Mar 13 & 138 & 5015 & 2.98 & 0.01 & 1.42 & --0.31 & 0.35 & --0.02 &  25 $\pm$ 4 \\
        HIP 71837 & 2021 Feb 22 & 165 & 5030 & 2.73 & --0.3 & 1.65 & --0.53 & 0.29 & --0.14 &  26 $\pm$ 4 \\
        HIP 71837 & 2021 Apr 25 & 152 & 4929 & 2.68 & --0.27 & 1.38 &--0.56 & 0.23 & --0.08 &  23 $\pm$ 3\\
        HIP 71837 & 2022 Mar 12 & 167 &  4941 & 2.86 & --0.27 & 1.37 & --0.50 & 0.29 & --0.01 &   23 $\pm$ 3\\
        BD+27 2057 & 2021 Jan 22 & 153  & 4873 & 2.34& --0.56 & 1.35 & --0.69 & --0.39 & --0.23 & 5 $\pm$ 1\\
		\hline
	\end{tabular}
    \end{center}
    Solar Abundance Scale from \citet{magg22} for our work. We do not re-normalize results reported from \citet{afsar12} since they derived differing solar abundances on a species by species basis using their own solar spectrum.
\end{table*}

We selected all RHB/RC and RC stars from \citet{afsar12} and compared their [C/Fe] and [N/Fe] abundances to our results. The results are shown in Fig. \ref{fig:cnomg_abund}. Most of the \citet{afsar12} RC targets are metal-rich, thin disk stars. We find agreement between those abundance ratios and our young, $\alpha$-rich thin disk RC stars. 

\begin{figure*}[t!]
\centering
	\includegraphics[width=7.in]{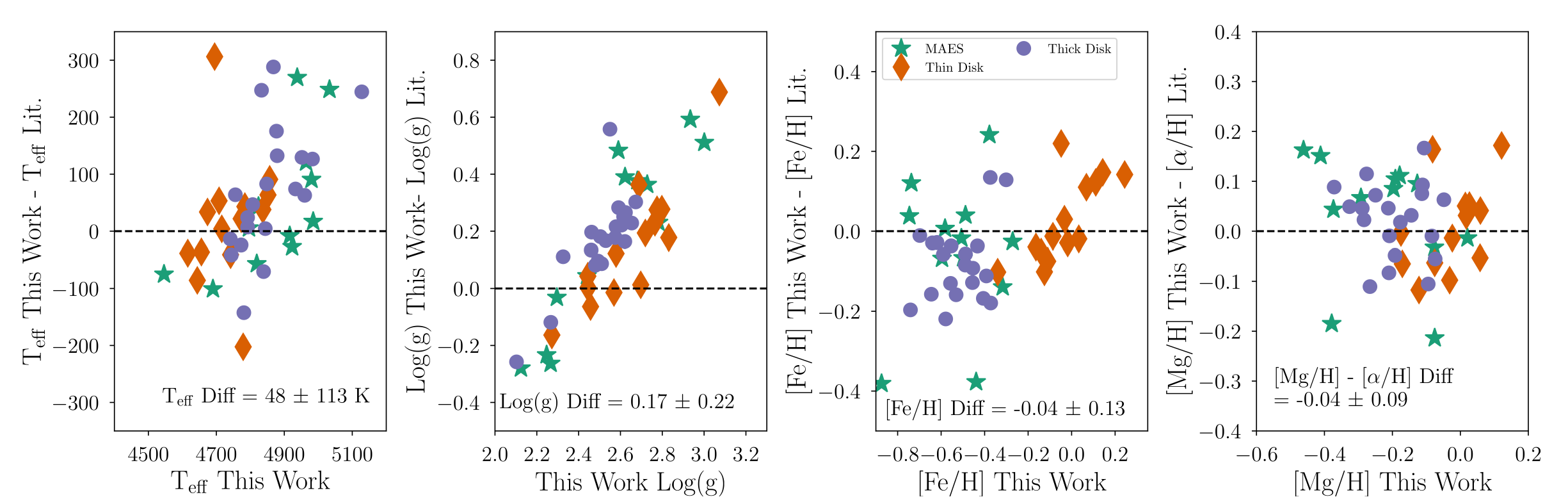}
    \caption{Atmospheric parameter comparison from Table \ref{tab:atmoparams} with LAMOST parameters from \citet{huang20}. The symbols are the same as Fig. \ref{fig:sample_abun_phot} and the dashed line represents no difference between the LAMOST value and our measurement.}
    \label{fig:param_comp}
\end{figure*}

\subsection{Comparison to LAMOST Parameters}

We compare the results from the BACCHUS pipeline with those from the RC catalog of \citet{huang20} in Fig \ref{fig:param_comp}. We do not measure the total $\alpha$-element abundance in our sample and so we compare our [Mg/H] abundance to their [$\alpha$/H] measurements as a consistency check on identifying $\alpha$-enriched stars. Our Fe and Mg abundances are normalized by the \citet{magg22} solar scale and the \citet{huang20} abundances are not altered. 

The average difference and standard deviation were calculated for each star in our sample and listed in Fig. \ref{fig:param_comp}. Typical uncertainties for the LAMOST values are 100 K, 0.1 dex, 0.1 dex, and 0.05 dex for \teff{}, log(g), [Fe/H], and [$\alpha$/H] \citep{luo15,xiang15,xiang17,huang20}. If we assume the total scatter is a combination of uncorrelated BACCHUS and LAMOST catalog errors, then the uncertainties for our sample would be 53 K, 0.2 dex, 0.08 dex, and 0.04 dex for \teff{}, log(g), [Fe/H], and [Mg/H].

Only our measured surface gravity is offset relative to the measurements from \citet{huang20} and has significant uncertainty. Our repeated measurements described in section \ref{sec:comp_to_afsar12} show that internal consistency for log(g) measurements are on the order of 0.1 - 0.2 dex, consistent with uncertainties estimated from the LAMOST comparisons. Additionally, the LAMOST log(g) values were derived using asteroseismic training samples which can have systematic differences from spectroscopically derived log(g) on the order of 0.1 - 0.2 dex depending on spectral type \citep{pinsonneault14,xiang17,lagarde24}. Similar changes in systematic offsets that change with log(g) and \teff{} have also been observed when comparing LAMOST values to APOGEE abundances \citep{xiang17,ho17}. Although the log(g) values are offset, the carbon isotope ratio is minimally sensitive to changes in atmospheric parameters. For example, \citet{lagarde24} finds a change of log(g) = 0.22 dex has an offset of $\Delta$\cratio{} = $\pm$ 1 using the CN line, \citet{tautvaisien13} finds varying the log(g) by +0.3 dex led to $\Delta$\cratio{} between --2 and 0 using the CN line, and \citet{maas19} found a change of 0.2 dex for log(g) created an average $\Delta$\cratio{} = 0.1 using CO lines. Changes with other atmospheric parameters are also found to offset the carbon isotope ratio by $\Delta$\cratio{} $\sim$ 1-2 \citep{lagarde24} 

\subsection{Other Literature Comparisons}

We compare our derived Mg and O abundances with literature FGK stars from \citet{bensby14} in Fig. \ref{fig:cnomg_abund}. We expect the [O/Fe] and [Mg/Fe] vs. [Fe/H] relationship to match known Milky Way chemical evolution patterns, while C and N may be uniquely affected by evolutionary state for RC stars. The abundances of \citet{bensby14} were placed on the abundance scale \citet{magg22}. We find broad agreement between our abundance ratios and those of \cite{bensby14}, further demonstrating the reliability of our abundance measurement methodology. 

Lastly, we compared our C and N abundances to red clump stars from \citet{afsar12,lagarde24} in Fig. \ref{fig:cnomg_abund}. Our carbon abundances are similar to the literature abundances for both stellar populations. For N, our  abundances are $\sim$ 0.1-0.3 dex higher for both stellar populations, with the most metal-poor \citet{lagarde24} stars are the most offset from our results. The abundances may differ due to different abundance analysis methodologies. For example, we fit all CN lines simultaneously and used the oxygen triplet to determine O abundances. \citet{lagarde24} used the [O I] 6300.31 \ang{} when available (and used Mg as a proxy for O when the [O I] feature was telluric contaminated) and fit seven strong CN features individually. 

\subsection{Final Uncertainties}

Final abundances and reported uncertainties in table \ref{tab:abund_appendix} were derived from the mean abundance and standard deviation on abundances from the set of lines available for each element. Two exceptions are the nitrogen uncertainty and the \cratio{} ratio uncertainty. The nitrogen abundance was fit from multiple lines simultaneously and carbon isotope ratio uncertainty is derived from repeated measurements.

We expect the SNR-statistical uncertainty to be the most significant source of uncertainty due to the weak $^{13}$CN line strengths. The carbon isotope ratio uncertainties were derived from a Monte Carlo simulation to best characterize the errors fitting the one absorption line, 2500 mock observations were generated for each star by treating each point in the observed spectrum as a Gaussian, with a mean value of the observed normalized flux and the standard deviation set by the signal-to-noise ratio of the data. The data was fit using the same $\chi^{2}$ minimization technique and the 16$\%$ and 84 $\%$ values were adopted as the 1$\sigma$ uncertainties reported in Table \ref{tab:abund_appendix}. 

\section{Discussion}
\label{sec:discussion}

\subsection{Star Kinematics and Li Abundance Estimates}
\label{subsec:li_kinematicss}

We calculate stellar population membership probabilities using UVW velocities since previous studies have used full 6-d phase space motion to distinguish between thin and thick disk stars. Kinematics can provide additional verification on population for our age-[$\alpha$/Fe] classification system, although chemical abundance may provide a cleaner sample selection than kinematics \citep{aguirre18}. Other studies find MAE stars generally have similar UVW velocities as thick disk stars \citep{sun20,zhang21} and our calculations can help distinguish between formation mechanism and provide evidence of potential sample contamination.  

We determined the probabilities using the same methodology as \citet{maas22}. To summarize, we used kinematic information from \citet{gaiadr3} with zero-point corrections applied to the parallax values using the calculation from \citet{lindegren21}. UVW velocities and errors were calculated using \texttt{pyia} \citep{price-whelan18}, coordinate transformations were done with \texttt{Astropy} \citep{astropy13}, and population probabilities from \citet{ramirez13}. Our final UVW velocities are listed in Table \ref{tab:abund_appendix}.

We find that the thin disk and thick disk memberships based on chemistry and age in Fig. \ref{fig:sample_abun_phot} are broadly consistent with population assignment based on kinematics. All of our thin disk stars have kinematic probabilities consistent with the chemistry/age selection. For thick disk stars, 10 stars have 50$\%$ or higher thick disk membership probability, five stars have a 90$\%$ or larger thin disk membership probability, and one thick disk star has a 61$\%$ probability of being a halo star. 

We also estimated the lithium abundances using the feature at 6707 \ang{} to aid in constraining formation mechanisms of the MAE stars. For example, high Li abundances may be a signal of mergers between red giants and white dwarfs, especially when combined with \cratio{} ratio measurements \citep{zhang20,maben23}. We used \texttt{TSFitPy} to estimate the LTE abundance for each star to determine if any star may be potentially Li-rich, with a threshold of A(Li) $>$ 1.5 dex (e.g. \citealt{aguileragomez23}). We found no enhanced Li abundance for any of our MAE stars and only one thin disk star met the criteria and with an estimated A(Li) = 1.82 dex (DR3 4009353280781776384).  

\subsection{Carbon Isotope Ratios for Red Clump Stars}

\begin{figure}
\centering
	\includegraphics[width=3.3in]{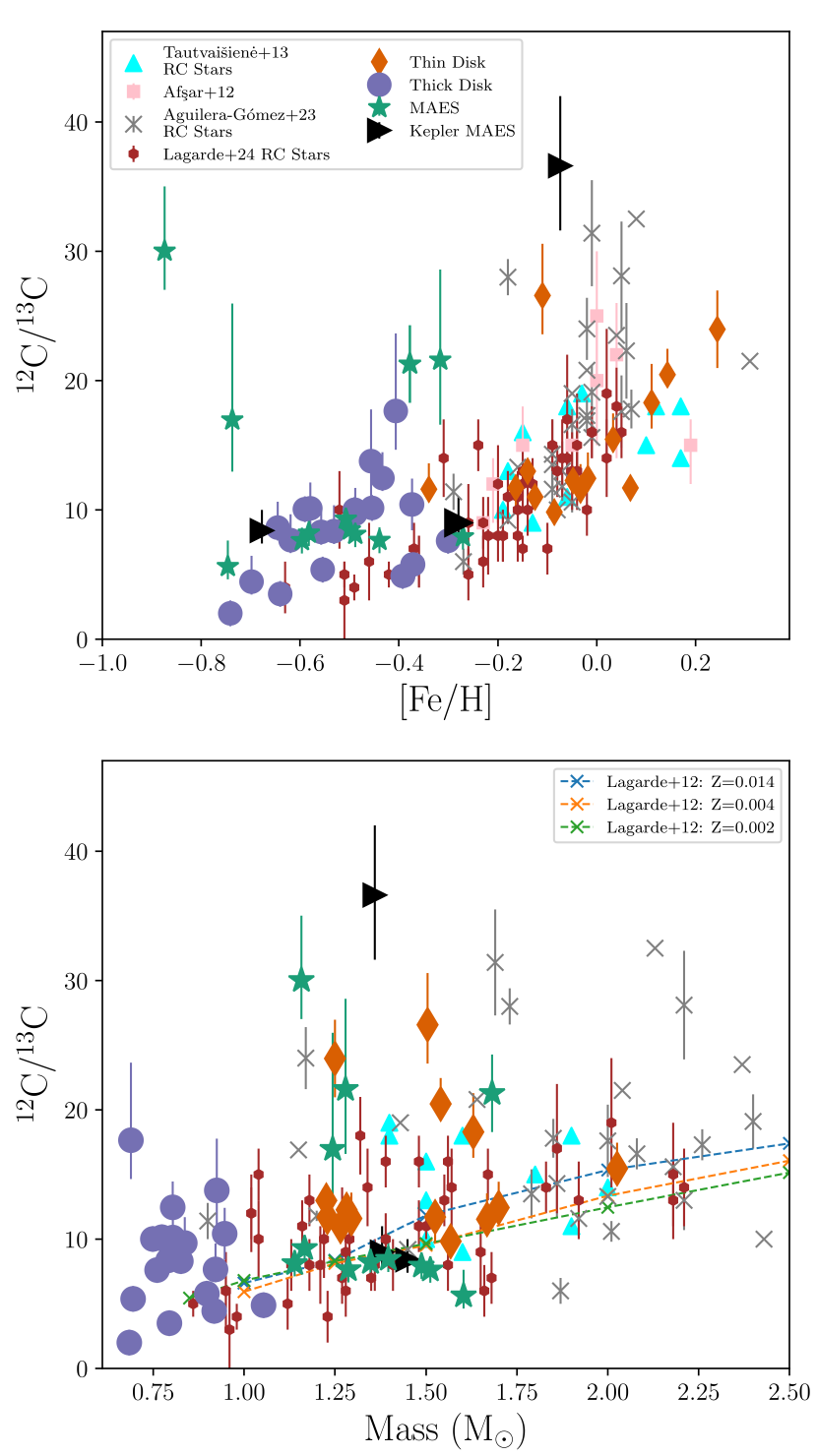}
    \caption{The carbon isotope ratios from our sample stars are plotted with reference to metallicity (left) and the stellar mass (right panel). Masses were adopted from \citet{sun20}. The symbols are the same as Fig. \ref{fig:cnomg_abund}.} 
    \label{fig:carboniso_rc_metallicity_mass}
\end{figure}

The \cratio{} as a function of [Fe/H] and mass are plotted in Fig. \ref{fig:carboniso_rc_metallicity_mass} for the thin disk, thick disk, and MAE star samples studied in this work. We increased the comparison size by adding literature abundance measurements from \citet{afsar12,aguileragomez23,lagarde24}. In all samples we chose only HB or RC stars, and stars with probabilities greater than 95$\%$ of being in the RC from \citet{aguileragomez23}. 

We find that our thin disk stars are generally consistent with other red clump stars near solar metallicities. The only thin disk star with a \cratio{} ratio above 20 is a mildly Li rich star with A(Li) $\sim$ 1.8 dex, as discussed in subsection \ref{subsec:li_kinematicss}. \citet{aguileragomez23} found RC stars with Li near $\sim$ 1.5 that also had high \cratio{} $>$ 20, in addition to RC stars with A(Li) $<$ 1 and low Li. Our thick disk sample also appears have decreasing \cratio{} ratios with decreasing metallicity, although there is significant scatter in the abundances. Overall, the average thin disk isotopic ratio is $<$\cratio{}$>$ = 14.8 $\pm$ 5.2 and $<$\cratio{}$>$ = 8.2 $\pm$ 3.4 for thick disk stars. Removing the one Li rich star, with a high carbon isotope ratio leads to an average of  $<$\cratio{}$>$ = 13.9 $\pm$ 4.2 for the thin disk. 
 
While the relationship between \cratio{} and mass has significant scatter, as shown in Fig. \ref{fig:carboniso_rc_metallicity_mass}, the average difference between thin disk and thick disk stars is similar to the models of \citet{lagarde12} that include thermohaline mixing. For Z = 0.014 (approximately solar metallicity), the 1 solar mass RC model has a \cratio{} = 6.6 and the model with a mass of 1.5 has a \cratio{} = 11.8. Our abundances that sample to lower metallicity evolved stars may help aid in constraining future models of extra-mixing for stars between 0.8 - 2 solar masses. 

The statistical methods used to measure star masses may introduce random uncertainties or systematic biases. Underestimated mass uncertainties may result in misclassified MAE stars that are likely low-mass thick disk stars. Potential stellar population contamination makes drawing conclusions on typical \cratio{} ratios for MAE stars potentially challenging. For systematic errors, a correlation with mass derived from KPCA analysis and [C/N] would bias our sample against finding stars from the multiple different pathways MAE stars might form. However, We do find that our MAE stars span a range of [C/N], similar to previous studies of MAE stars \citep{hekker19}.

\subsection{MAES RC Stars with Anomalous \cratio{} Ratios: Possible Formation Pathways}
\label{subsec:anomalous_rc}

10 out of our 15 red clump MAE stars have carbon isotope ratios consistent with thick disk stars at similar metallicities. Our results are in-line with other studies of thick disk stars that find carbon abundances, kinematics, and $\alpha$-enrichment for MAE stars are similar to thick disk members \citep{sun20,zhang21,cerqui23}. However, a small number of our MAE stars have \cratio{} ratios significantly higher than other red clump stars at similar metallicities. Five stars in total have \cratio{} $>$ 15 and as seen in Fig. \ref{fig:carboniso_rc_metallicity_mass}; four RC stars selected from \citet{sun20,huang20} and one star with a mass identified from asteroseismology. While these stars have different metallicities, none have enhanced Li ratios making some merger scenarios, such as a merger with a white dwarf, unlikely \citep{zhang20}. The properties of the outlier stars are described below. 

\textbf{KIC 3833399:} KIC 3833399 is the high metallicity, carbon isotope ratio outlier, with a mass of 1.48 M$_{\odot}$ \citep{jofre23}. The carbon isotope ratio is approximately 4$\sigma$ away from our average carbon isotope ratio of 14.8 for thin disk stars although this star has a 78$\%$ chance of belonging to the thin disk population based on kinematics. Additionally, long-term radial velocity monitoring from \citet{jofre23} found no evidence that KIC 3833399 had a binary companion. A stellar merger scenario must have increased the total mass and increased the surface \cratio{} ratio while not enhancing the total [C/Fe] abundance which may be expected from mass-transfer with an AGB companion. Another possibility is that this massive star is genuinely young as evidenced from the high \cratio{} and kinematic classification; a scenario suggested for other MAE stars \citep{lu24}.  

\textbf{Gaia DR3 672933143771515392:} Gaia DR3 672933143771515392 is the most metal poor star in the sample with an [Fe/H] = --0.87, a \cratio{} ratio of 30 $\pm$ 5, and a 89$\%$ probability of being a thick disk member. The star also has high carbon and nitrogen abundances with [C/Fe] = 0.28 dex and [N/Fe] = 1.00 compared to other MAE stars. This star was also identified as a CH star from LAMOST spectra with enhanced CH lines in the G band, CN at 4215 \ang{}, and Ba II features \citep{wei16}. Mass-transfer with an AGB star companion may explain these abundance patterns since AGB stars have enhanced  $^{12}$C and N abundances \citep{smith90}. Gaia DR3  672933143771515392 appears to be an RC star that previously had an AGB star companion and additional s-process element abundance ratios may further constrain the mass of the potential companion. 

\textbf{Gaia DR3 638939267896892800:} Gaia DR3 638939267896892800 is both kinematically similar to thin disk stars and has a \cratio{} = 21 $\pm$ 3 which is $\sim$ 2$\sigma$ away from the average thin disk carbon isotope ratio. LAMOST spectra also identified this star as a CH star \citep{wei16}. \citet{shejeelammal21} finds a lower metallicity [Fe/H] = --0.53 and a higher carbon abundance of [C/Fe] = 0.23 (after adjusting the solar normalization to match our value of A(C)${|_\odot}$ = 8.56). Our carbon abundance of [C/Fe] = 0.12 is enhanced relative to other RC stars and consistent within the uncertainties with \citet{shejeelammal21}. Overall, this star has CNO abundances consistent with mass transfer with an AGB companion. 

\textbf{Gaia DR3 3671355056986858880}: Gaia DR3 3671355056986858880 was measured to have an [Fe/H] = --0.32 and a \cratio{} = 22$^{+7}_{-5}$. The \cratio{} ratio is 2.75$\sigma$ higher than the average thick disk carbon isotope ratio. Unlike the previous anomalous stars, this object is not well-studied in the literature. We do not find an increased carbon abundance ([C/Fe] = --0.05) suggesting mass-transfer with an AGB companion is not as likely as other stars with high \cratio{} ratios in our sample. Data from Gaia DR3 finds no obvious sign of binarity; the star has a RUWE of 1.05 and a Gaia DR3 radial velocity error of 0.26 km s$^{-1}$ \citep{gaiadr3}. The kinematic analysis suggests that this star is a likely thick disk member and the binary interaction that created the mass enhancement must have happened in a manner that allowed the carbon isotope ratio to be consistent with a higher mass star. 

The abundances are also consistent with a sub-giant star that is undergoing the first dredge-up. The potential the star is misclassified is unlikely since both the atmospheric parameters and photometry in Fig. \ref{fig:sample_abun_phot} show that Gaia DR3 3671355056986858880 are consistent with an RC star. Alternatively, the star may be heated by other temporary properties that would make a sub-giant star bluer and more luminous. Temporary tidal heating may affect stars in open cluster \citep{arthur24} can cause a similar effect. However, tidal heating would move energy from the orbit and cause stellar mergers. The high kinematic probability of being a thick disk star would suggest that we are not catching the star in a rare instance of tidal heating and longer-term radial velocity monitoring and s-process abundance measurements would shed light on the origins of this star. Large changes to the inner structure happen when two stars merge. For example, a star created through a stellar merger is expected to have over-massive envelopes relative to their core masses compared to similar mass single stars \citep{rui21}. The high \cratio{} ratio may aid in constraining models of stars created through full merger scenarios.  

\textbf{Gaia DR3 971925848973504512}: Gaia DR3 971925848973504512 has a \cratio{} = 17$^{+9}_{-4}$ which is approximately 2$\sigma$ higher than other stars at similar metallicities near [Fe/H] = --0.74. However, the star is otherwise similar to low-mass thick disk stars chemically and kinematically. This star has no large enhancement of carbon or nitrogen and has a 93$\%$ probability of being a thick disk star. The star has a RUWE of 0.96 and no large RV uncertainties from Gaia DR3 \citep{gaiadr3} suggesting the object is not in a binary system. The isotope ratio seems consistent for a single-star evolution of a 1.5 M$_{\odot}$. This star may have undergone a stellar merger before either object evolved off the main sequence.  Additionally, the measurement is only 2$\sigma$ from similar thick disk stars and may be explained by statistical uncertainties within the carbon isotope ratio measurement. 

\subsection{Carbon Isotope Ratios for Red Giant Stars}
\label{subsec:red_giant}
\begin{figure}
	\includegraphics[width=3.3in]{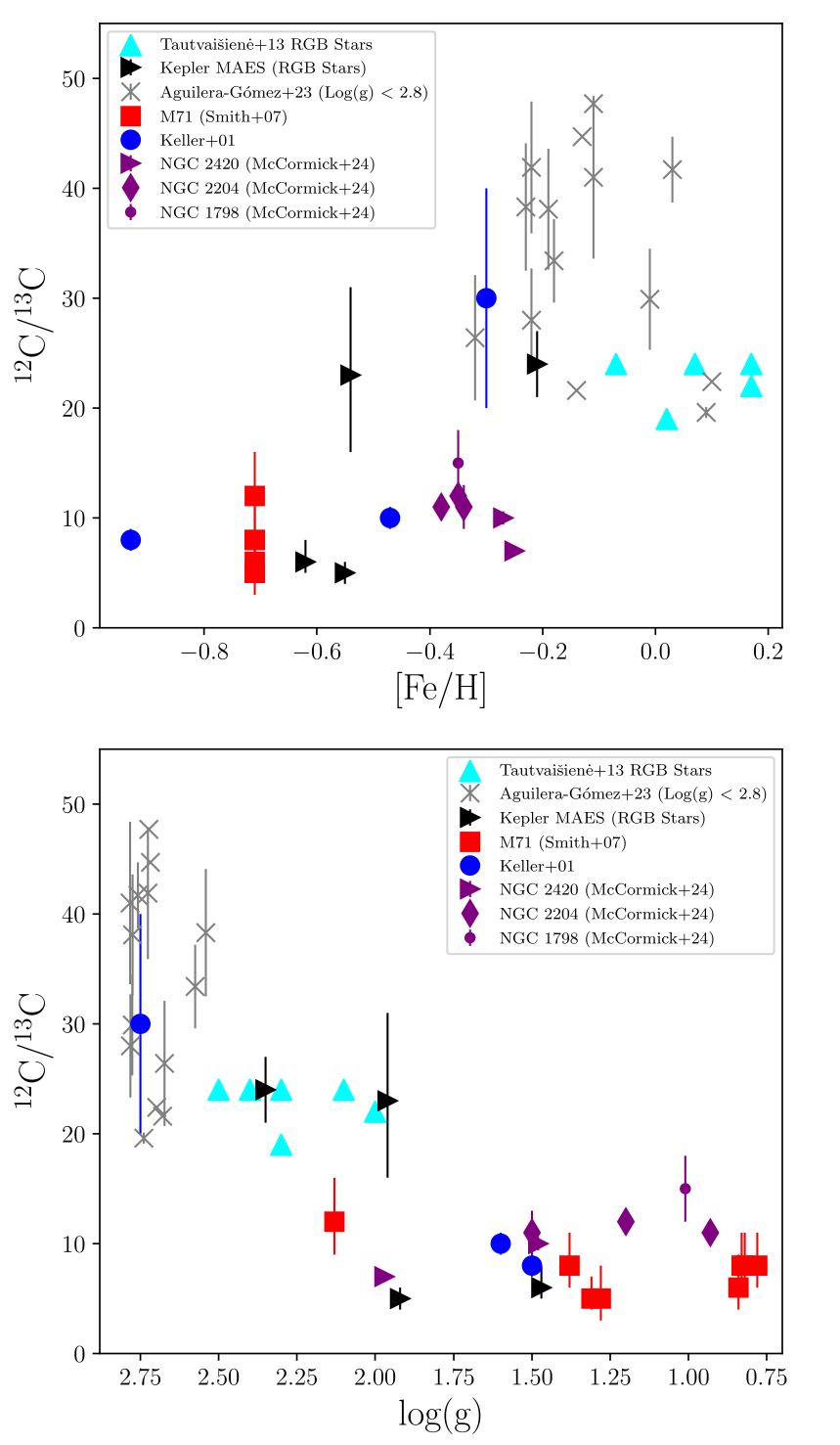}
    \centering
    \caption{The carbon isotope ratios from our sample stars are plotted with reference to metallicity (left) and the surface gravity (right panel).  The symbols are the same as Fig. \ref{fig:cnomg_abund}. The solid points are stars within the RC photometric sample (shown in Fig. \ref{fig:sample_abun_phot}).} 
    \label{fig:carboniso_rgb_metallicity_logg}
\end{figure}

We also observed a small sample of stars identified as MAES stars on the RG branch. Since the \cratio{} ratio changes as a red giant evolves, we cannot directly compare the surface composition of these objects to our RC sample. Specifically, the carbon isotope ratio depends strongly on position on the red giant branch and on initial metallicity. The first dredge up initially lowers the \cratio{} to values near 30-40 and additional mixing occurs after the LF bump, decreases the value further. We therefore compare our results to literature sources of open clusters \citep{mccormick23}, globular clusters \citep{smith07}, and disk stars \citep{keller01,tautvaisien13,aguileragomez23} in Fig. \ref{fig:carboniso_rgb_metallicity_logg}. 

We find that three of our Kepler MAE stars have carbon isotope ratios similar to stars at similar metallicities and evolutionary states. However, the star KIC5966873 has a high \cratio{}, that is comparable to high-metallicity open cluster members of \citet{tautvaisien13}. The high carbon isotopic ratio suggests that extra-mixing has not yet begun in KIC 5966873 or that there has been mass-transfer of  $^{12}$C-rich material. This star does not have an unusual CNO abundances relative to the other Kepler red giant stars, suggesting that the source of the mass-transfer from a carbon rich object is unlikely. Our limited sample inhibits definitive descriptions on the formation pathway of this object but additional carbon isotope ratio measurements of red giant stars will shed light on if this star is truly an outlier. 

\section{Conclusions}
\label{sec:conclusion}

We explored the origins of MAE stars by measuring the carbon isotope ratios for red clump stars identified from LAMOST data \citep{huang20,sun20} and in stars with asteroseismic masses determined from Kepler observations. Abundance analysis was conducted with Turbospectrum using both the TSFitpy and BACCHUS codes. Overall, we measured metallicities, CNO abundances, and carbon isotope ratios in 19 MAE stars (15 red clump stars and four red giant stars), 14 thin disk stars, and 20 thick disk stars. 10 of the red clump MAE stars had similar carbon isotope ratios, between 6 $<$ \cratio{} $<$ 8, to thick disk stars at similar metallicities. Carbon isotope ratios were not found near the CNO burning equilibrium of 3.5 that is observed when significant amount of material is dredged up for metal-poor red giants beyond the luminosity function bump. Additionally, some of the MAE stars have similar isotope ratios seen in single low mass stars and five of the red clump MAE stars have high \cratio{} ratios between 17 - 37. These stars span a metallicity range of -- 0.87 $<$ [Fe/H] $<$ --0.07. Two of the anomalous stars appear have [C/Fe] and [N/Fe] abundance ratios consistent with mass-transfer from an AGB star companion. The other three stars have no signatures of binarity and the carbon isotope ratio may constraints on potential merger scenarios. 

We find three MAE red giant stars have  carbon isotope ratios consistent with other giants studied in the literature at similar metallicities and surface gravities. However, one red giant, KIC 5966873, has a \cratio{} ratio of 23$^{+8}_{-7}$, around $\sim$ 2$\sigma$ larger than similar red giants. The carbon isotope ratio better resembles high metallicity red giant stars and the \cratio{} ratio may constrain the formation mechanism of this MAE star. Overall, the carbon isotope ratio may provide a unique constraint on the evolutionary history of MAE stars and distinguish between different formation pathways. 

\begin{acknowledgments}
We thank the referee for their thoughtful feedback and suggestions that improved this manuscript. This work has made use of data from the European Space Agency (ESA) mission {\it Gaia} (\url{https://www.cosmos.esa.int/gaia}), processed by the {\it Gaia} Data Processing and Analysis Consortium (DPAC, \url{https://www.cosmos.esa.int/web/gaia/dpac/consortium}). Funding for the DPAC has been provided by national institutions, in particular the institutions participating in the {\it Gaia} Multilateral Agreement. This paper includes data taken at The McDonald Observatory of The University of Texas at Austin. This research has made use of the SIMBAD database, operated at CDS, Strasbourg, France. This research has made use of the VizieR catalogue access tool, CDS, Strasbourg Astronomical Observatory, France (DOI : 10.26093/cds/vizier). We thank Gloria Koenigsberger for excellent discussions on this research that improved the manuscript. ZGM is partially supported by a NASA ROSES-2020 Exoplanet Research Program Grant (20-XRP20 2-0125). CM is supported by the NSF Astronomy and Astrophysics Fellowship award number AST-2401638. KH is partially supported through the NSF Astronomy and Astrophysics Grants AST-2407975 and AST-2108736 along with the Wootton Center for Astrophysical Plasma Properties funded under the United States Department of Energy collaborative agreement DE-NA0003843. 
\end{acknowledgments}

\facilities{McDonald Observatory: 2.7 m Harlan J. Smith
Telescope (Tull Coud\'e)}

\software{\texttt{scipy} \citep{scipy20}, \texttt{numpy} \citep{numpy}, \texttt{matplotlib} \citep{hunter07}, \texttt{astropy} \citep{astropy13,astropy18,astropy22}, \texttt{pyia} \citep{price-whelan18}, \texttt{dustmaps} \citep{green19}, \texttt{iSpec} \citep{blanco-cauresma14,blanco-cauresma19}, \texttt{Turbospectrum} \citep{plez12}, and \texttt{TSFitPy} \citep{gerber22,storm23}}

\appendix

\section{Appendix 1: Star Information}

The full abundance information for our sample is listed in Table \ref{tab:abund_appendix} in this appendix. 

\begin{longrotatetable}
\startlongtable
\centerwidetable
\begin{deluxetable*}{lcccccccccccccccccccccc} 
\tablecaption{Abundances and Kinematics for Sample Stars\label{tab:abund_appendix}} 
\tablewidth{0pt} 
\centering
\tabletypesize{\tiny}
 \tablehead{\colhead{Gaia DR3 ID} & \colhead{[Fe/H]} &  \colhead{[C/Fe]} & \colhead{$\sigma$ [C/Fe]} &  \colhead{[N/Fe]} & \colhead{[O/Fe]} &  \colhead{$\sigma$[O/Fe]} &  \colhead{[Mg/Fe] } &  \colhead{$\sigma$[Mg/Fe]}  & \colhead{$\mathrm{\frac{12C}{13C}}$}  & \colhead{$\sigma$ $\mathrm{\frac{12C}{13C}}$} & \colhead{$\sigma$ $\mathrm{\frac{12C}{13C}}$} & \colhead{U} & \colhead{V} & \colhead{W}  & \colhead{$\sigma$U} & \colhead{$\sigma$V} & \colhead{$\sigma$W} & \colhead{P(Thin)} & \colhead{P(Thick)} & \colhead{P(Halo)}\\  \colhead{} & \colhead{(dex)} & \colhead{(dex)}& \colhead{(dex)}& \colhead{(dex)}& \colhead{(dex)}& \colhead{(dex)}& \colhead{(dex)}& \colhead{(dex)} & \colhead{} & \colhead{High} & \colhead{Low} & \colhead{(kms$^{-1}$)} & \colhead{(kms$^{-1}$)} & \colhead{(kms$^{-1}$)} & \colhead{(kms$^{-1}$)}& \colhead{(kms$^{-1}$)}& \colhead{(kms$^{-1}$)}& \colhead{}& \colhead{}& \colhead{}}
\startdata 
  \multicolumn{21}{c}{Thin Disk}  \\ 
        \hline \\
866507457241940480 & -0.02 & -0.31 & 0.11 & 0.22 & 0.02 & 0.01 & -0.01 & 0.12 & 12 & 2 & 1 & -29.35 & -22.75 & 14.28 & 0.14 & 0.18 & 0.11 & 0.98 & 0.02 & 0.00 \\
688166190179213184 & 0.11 & -0.26 & 0.15 & 0.72 & 0.21 & 0.14 & -0.08 & 0.12 & 18 & 3 & 2 & 12.96 & 8.42 & 0.12 & 0.13 & 0.06 & 0.12 & 0.99 & 0.01 & 0.00 \\
3138549726673657600 & 0.03 & -0.31 & 0.12 & 0.46 & 0.05 & 0.10 & -0.06 & 0.09 & 15 & 2 & 1 & -12.42 & 31.03 & 9.71 & 0.25 & 0.33 & 0.06 & 0.98 & 0.02 & 0.00 \\
4009353280781776384 & -0.11 & -0.21 & 0.10 & 0.27 & 0.07 & 0.06 & 0.03 & 0.04 & 27 & 4 & 3 & 22.25 & -16.45 & -11.52 & 0.11 & 0.33 & 0.16 & 0.99 & 0.01 & 0.00 \\
1312079562236150656 & -0.03 & -0.33 & 0.14 & 0.36 & 0.03 & 0.01 & 0.05 & 0.07 & 12 & 2 & 1 & 28.19 & 12.23 & -24.08 & 0.16 & 0.12 & 0.21 & 0.98 & 0.02 & 0.00 \\
1388620891675799168 & -0.16 & -0.29 & 0.09 & 0.27 & 0.07 & 0.06 & -0.01 & 0.10 & 12 & 1 & 2 & -20.95 & -18.75 & -14.85 & 0.12 & 0.07 & 0.11 & 0.99 & 0.01 & 0.00 \\
3833688568619253120 & 0.14 & -0.28 & 0.13 & 0.67 & 0.12 & 0.15 & -0.08 & 0.13 & 20 & 2 & 1 & -5.80 & -19.31 & 3.67 & 0.12 & 0.14 & 0.18 & 0.99 & 0.01 & 0.00 \\
842398878096774784 & -0.05 & -0.26 & 0.11 & -0.08 & -0.28 & 0.29 & -0.03 & 0.11 & 12 & 1 & 1 & 22.34 & -32.75 & -22.11 & 0.11 & 0.28 & 0.12 & 0.96 & 0.04 & 0.00 \\
1330490025449104640 & -0.12 & -0.23 & 0.13 & 0.35 & 0.13 & 0.08 & 0.00 & 0.07 & 11 & 1 & 1 & -46.64 & 35.82 & -0.81 & 0.31 & 0.21 & 0.10 & 0.97 & 0.03 & 0.00 \\
3326023296796724096 & 0.24 & -0.21 & 0.11 & 0.61 & 0.14 & 0.16 & -0.19 & 0.06 & 24 & 3 & 3 & -1.69 & -31.29 & -0.61 & 0.18 & 0.28 & 0.10 & 0.98 & 0.02 & 0.00 \\
3976449776845278592 & -0.09 & -0.27 & 0.12 & 0.42 & 0.09 & 0.09 & 0.11 & 0.04 & 10 & 1 & 1 & -21.77 & -53.62 & -1.22 & 1.14 & 1.84 & 1.00 & 0.88 & 0.12 & 0.00 \\
3976085529258764800 & 0.07 & -0.18 & 0.10 & -0.08 & -0.26 & 0.46 & -0.05 & 0.07 & 12 & 1 & 1 & -24.36 & -10.48 & -9.29 & 0.36 & 0.18 & 0.18 & 0.99 & 0.01 & 0.00 \\
1486503509885506944 & -0.34 & -0.27 & 0.10 & 0.39 & 0.14 & 0.04 & 0.17 & 0.11 & 12 & 2 & 1 & 70.26 & -4.29 & 29.09 & 0.49 & 0.22 & 0.25 & 0.96 & 0.04 & 0.00 \\
1447162502806921472 & -0.14 & -0.24 & 0.12 & 0.34 & 0.10 & 0.05 & 0.26 & 0.21 & 13 & 1 & 1 & 13.08 & 11.73 & -20.89 & 0.08 & 0.07 & 0.12 & 0.99 & 0.01 & 0.00 \\
\\  \hline
  \multicolumn{21}{c}{Thick Disk}  \\ 
        \hline \\
3130104824535988096 & -0.49 & -0.13 & 0.14 & 0.31 & 0.41 & 0.02 & 0.34 & 0.11 & 10 & 2 & 1 & 2 & -57.34 & -46.81 & 0.48 & 0.62 & 0.69 & 0.41 & 0.58 & 0.01 \\
3222153253655970176 & -0.55 & -0.04 & 0.05 & 0.34 & 0.37 & 0.00 & 0.27 & 0.06 & 5.00 & 1 & 1 & 54.03 & -31.23 & 10.05 & 0.29 & 0.58 & 0.10 & 0.96 & 0.04 & 0.00 \\
1544130323049089536 & -0.39 & -0.06 & 0.15 & -0.25 & 0.32 & 0.06 & 0.31 & 0.11 & 5 & 1 & 1 & 23.68 & -70.85 & 25.62 & 0.13 & 1.00 & 0.40 & 0.33 & 0.66 & 0.01 \\
723448831119162752 & -0.59 & -0.01 & 0.03 & 0.39 & 0.44 & 0.02 & 0.34 & 0.13 & 10 & 1 & 1 & 1.19 & -52.37 & -35.37 & 3.02 & 2.91 & 6.61 & 0.74 & 0.26 & 0.00 \\
3096606828405802240 & -0.41 & -0.01 & 0.03 & 0.44 & 0.53 & 0.07 & 0.33 & 0.13 & 18 & 6 & 3 & 21.25 & 12.23 & -29.09 & 0.18 & 0.13 & 0.45 & 0.98 & 0.02 & 0.00 \\
3811610203335180416 & -0.64 & -0.01 & 0.05 & 0.30 & 0.31 & 0.04 & 0.36 & 0.06 & 4  & 1 & 1 & 64.44 & -75.80 & 21.05 & 0.78 & 0.78 & 0.39 & 0.13 & 0.86 & 0.02 \\
1295525109008882816 & -0.37 & -0.15 & 0.13 & 0.09 & 0.32 & 0.03 & 0.28 & 0.07 & 6 & 1 & 1 & -47.18 & 0.71 & -31.13 & 0.68 & 0.27 & 0.21 & 0.97 & 0.03 & 0.00 \\
3941980190394162560 & -0.45 & -0.07 & 0.10 & 0.39 & 0.44 & 0.07 & 0.26 & 0.08 & 10 & 1 & 1 & 201.57 & -68.33 & -62.93 & 4.26 & 1.79 & 0.59 & 0.00 & 0.39 & 0.61 \\
3869545635666293504 & -0.43 & -0.10 & 0.11 & 0.33 & 0.33 & 0.05 & 0.26 & 0.14 & 12 & 2 & 1 & -7.55 & -101.29 & -30.84 & 0.25 & 1.63 & 1.06 & 0.00 & 0.96 & 0.03 \\
3283445774160853632 & -0.74 & 0.01 & 0.02 & 0.56 & 0.58 & 0.02 & 0.45 & 0.03 & 2 & 1 & 1 & -63.73 & -83.45 & 7.21 & 0.21 & 0.96 & 0.48 & 0.06 & 0.92 & 0.02 \\
1244397371640334080 & -0.53 & -0.03 & 0.04 & 0.37 & 0.43 & 0.06 & 0.26 & 0.10 & 8 & 2 & 1 & -23.07 & -36.04 & 7.91 & 0.48 & 0.64 & 0.24 & 0.97 & 0.03 & 0.00 \\
640282939825239936 & -0.56 & -0.03 & 0.05 & 0.38 & 0.46 & 0.03 & 0.35 & 0.13 & 8 & 2 & 1 & 13.44 & -70.03 & -10.41 & 0.12 & 0.81 & 0.34 & 0.50 & 0.50 & 0.00 \\
4008867227922925824 & -0.62 & -0.01 & 0.02 & 0.50 & 0.45 & 0.05 & 0.29 & 0.10 & 8 & 2 & 1 & -6.99 & -75.77 & 27.94 & 0.18 & 0.82 & 0.18 & 0.20 & 0.79 & 0.01 \\
3882249324293741440 & -0.58 & -0.01 & 0.06 & 0.31 & 0.50 & 0.05 & 0.37 & 0.10 & 10 & 2 & 2 & 59.35 & -34.19 & 37.38 & 1.07 & 0.54 & 0.23 & 0.85 & 0.15 & 0.00 \\
3723876425343292032 & -0.70 & 0.01 & 0.04 & -0.20 & 0.12 & 0.01 & 0.33 & 0.07 & 4 & 2 & 1 & -107.23 & -71.60 & 48.92 & 1.99 & 1.34 & 0.72 & 0.01 & 0.94 & 0.05 \\
3973325544619220224 & -0.65 & -0.05 & 0.12 & 0.37 & 0.52 & 0.04 & 0.43 & 0.10 & 9 & 2 & 1 & 49.83 & -153.31 & -5.69 & 0.71 & 2.43 & 0.74 & 0.00 & 0.61 & 0.39 \\
3896192196887050368 & -0.30 & -0.21 & 0.20 & 0.78 & 0.52 & 0.09 & 0.19 & 0.14 & 8 & 1 & 1 & -28.18 & -45.50 & -31.99 & 0.84 & 1.14 & 0.61 & 0.85 & 0.15 & 0.00 \\
3163974593033517824 & -0.46 & -0.13 & 0.08 & 0.22 & 0.21 & 0.02 & 0.41 & 0.05 & 14 & 4 & 3 & -70.37 & -37.08 & -75.67 & 0.48 & 0.25 & 1.38 & 0.06 & 0.90 & 0.04 \\
3813076676968705408 & -0.49 & -0.08 & 0.10 & 0.31 & 0.40 & 0.03 & 0.38 & 0.13 & 10 & 1 & 1 & -82.83 & -37.09 & 45.77 & 1.22 & 0.28 & 0.43 & 0.54 & 0.45 & 0.01 \\
1275573508452428288 & -0.37 & -0.10 & 0.10 & 0.31 & 0.43 & 0.04 & 0.27 & 0.09 & 10 & 2 & 2 & 62.26 & 6.23 & 8.16 & 0.59 & 0.13 & 0.24 & 0.98 & 0.02 & 0.00 \\
\\ \hline 
  \multicolumn{21}{c}{MAES}  \\ 
        \hline \\
672933143771515392 & -0.87 & 0.28 & 0.07 & 1.00 & 0.63 & 0.06 & 0.50 & 0.11 & 30 & 5 & 3 & -29.14 & -61.32 & -60.73 & 0.35 & 1.01 & 1.40 & 0.09 & 0.89 & 0.02 \\
638939267896892800 & -0.38 & 0.12 & 0.07 & 0.77 & 0.30 & 0.16 & 0.00 & 0.15 & 21 & 3 & 3 & -10.46 & -2.62 & 6.61 & 0.35 & 0.13 & 0.38 & 0.99 & 0.01 & 0.00 \\
607150428232731008 & -0.51 & 0.02 & 0.03 & 0.36 & 0.55 & 0.14 & 0.31 & 0.16 & 9 & 1 & 1 & -51.78 & -80.73 & 8.54 & 0.13 & 0.40 & 0.40 & 0.12 & 0.87 & 0.01 \\
579491624824868608 & -0.27 & -0.19 & 0.16 & 0.84 & 0.51 & 0.09 & 0.19 & 0.14 & 8 & 1 & 1 & -0.26 & -20.40 & -3.51 & 0.72 & 1.13 & 1.45 & 0.99 & 0.01 & 0.00 \\
1235499505072865792 & -0.44 & -0.21 & 0.13 & 0.30 & 0.34 & 0.02 & 0.36 & 0.08 & 8 & 1 & 1 & 34.81 & -55.49 & -15.08 & 0.27 & 0.53 & 0.18 & 0.80 & 0.19 & 0.00 \\
971925848973504512 & -0.74 & -0.06 & 0.16 & 0.33 & 0.46 & 0.01 & 0.28 & 0.12 & 17 & 9 & 4 & 0.61 & -89.50 & 11.90 & 0.34 & 0.97 & 0.09 & 0.05 & 0.93 & 0.02 \\
3872367429179326592 & -0.58 & -0.28 & 0.22 & 0.72 & 0.47 & 0.04 & 0.29 & 0.09 & 8 & 1 & 1 & 4.03 & -114.72 & -83.23 & 0.39 & 2.96 & 1.72 & 0.00 & 0.76 & 0.24 \\
636614659798851712 & -0.60 & -0.32 & 0.25 & 0.79 & 0.45 & 0.03 & 0.42 & 0.08 & 8 & 1 & 1 & -122.42 & -54.48 & -91.04 & 1.61 & 0.68 & 2.20 & 0.00 & 0.77 & 0.23 \\
685849244302001408 & -0.49 & -0.28 & 0.16 & 0.57 & 0.42 & 0.07 & 0.30 & 0.06 & 8 & 1 & 1 & 118.26 & -47.90 & -11.66 & 0.82 & 1.16 & 0.34 & 0.41 & 0.57 & 0.02 \\
3876280792926214272 & -0.50 & -0.19 & 0.16 & 0.27 & 0.15 & 0.03 & 0.37 & 0.06 & 8 & 2 & 1 & 42.27 & -46.60 & -38.88 & 0.35 & 1.00 & 0.53 & 0.73 & 0.27 & 0.00 \\
3878388801529524480 & -0.75 & -0.11 & 0.22 & 0.58 & 0.52 & 0.07 & 0.33 & 0.12 & 6 & 2 & 1 & 15.86 & -29.34 & 70.66 & 0.86 & 0.18 & 0.48 & 0.34 & 0.64 & 0.01 \\
3671355056986858880 & -0.32 & -0.05 & 0.08 & 0.69 & 0.60 & 0.07 & 0.34 & 0.15 & 22 & 7 & 5 & -76.01 & -84.87 & 62.72 & 1.74 & 1.61 & 0.70 & 0.00 & 0.93 & 0.07 \\
 \\ \hline
  \multicolumn{21}{c}{Kepler Red Clump MAES }  \\ 
        \hline \\
2100234510918046592 & -0.07 & -0.08 & 0.09 & 0.60 & 0.37 & 0.06 & 0.19 & 0.07 & 37 & 6 & 5 & -8.24	&	-56.70	&	21.89	&	0.05	&	0.13	&	0.19	&	0.78	&	0.21	&	0.00 \\
2117411734401094528 & -0.68 & -0.09 & 0.20 & 0.51 & 0.49 & 0.02 & 0.49 & 0.16 & 8 & 2 & 1 & 102.90	&	-19.10	&	0.19	&	0.71	&	0.26	&	0.09	&	0.91	&	0.09 & 0.00	\\
2130894220860474752 & -0.23 & -0.13 & 0.04 & 0.44 & 0.39 & 0.10 & \nodata & \nodata & 9 & 2 & 1 & 70.70	&	-37.82	&	-34.81	&	0.79	&	0.42	&	0.37	&	0.79	&	0.21	&	0.00 \\
 \\ \hline
  \multicolumn{21}{c}{Kepler Red Giant MAES }  \\ 
        \hline \\
2077396108227657344  & -0.16 & -0.08 & 0.08 & 0.36 & 0.27 & 0.05 & 0.01 & 0.04 & 24 & 3 & 3 & 3.53	&	-37.11	&	-3.15	&	0.13	&	0.29	&	0.10	&	0.97	&	0.03	&	0.00 \\
2100961185027552384 & -0.50 & -0.37 & 0.12 & 0.60 & 0.41 & 0.20 & 0.15 & 0.05 & 5 & 1 & 1 & 24.52	&	-0.43	&	54.76	&	0.20	&	0.29	&	0.82	&	0.88	&	0.12	&	0.00\\
2103822354798961280 & -0.57 & -0.12 & 0.14 & 0.50 & 0.73 & 0.25 & 0.15 & 0.10 & 6 & 2 & 1 & 22.65	&	9.50	&	-43.56	&	0.25	&	0.24	&	0.86	&	0.95	&	0.05	&	0.00 \\
2077396108227657344 & -0.49 & -0.19 & 0.13 & 0.56 & 0.53 & 0.22 & 0.12 & 0.06 & 23 & 8 & 7 &	-20.98	&	-64.83	&	8.79	&	0.15	&	0.19	&	0.15	&	0.65	&	0.34	&	0.00\\
\enddata
\tablecomments{Note: Solar Abundance Scale from \citet{magg22}}
\end{deluxetable*}
\end{longrotatetable}

\bibliography{carbon_references}{}
\bibliographystyle{aasjournalv7}

\end{document}